\documentclass[aps,prl,showpacs,twocolumn]{revtex4}

\usepackage{epsfig}
\usepackage{longtable}

\begin{document}

\title{Identification of Extra Neutral Gauge Bosons at the LHC Using $b$- and $t$-Quarks}
\author{ Stephen Godfrey\footnote{Email: godfrey@physics.carleton.ca} 
and Travis A. W. Martin\footnote{Email: tmartin@physics.carleton.ca}}
\affiliation{
Ottawa-Carleton Institute for Physics, 
Department of Physics, Carleton University, Ottawa, Canada K1S 5B6 }

\date{\today}

\begin{abstract}
New Neutral Gauge Bosons, $Z'$'s, are predicted by many models of physics beyond the Standard 
Electroweak Theory.  It is possible that a $Z'$ would be discovered early in the Large
Hadron Collider program.  The next step would be to measure its properties
to identify the underlying
theory that gave rise to the $Z'$.  Heavy quarks have the unique property that they can be
identified in the final states. 
In this letter we demonstrate that measuring $Z'$ decays
to $b$- and $t$-quark final states can act as an effective means of discriminating between
models with extra gauge bosons.  
\end{abstract}
\pacs{14.70.Pw, 12.60.Cn, 12.15.Mm}

\maketitle


In the coming years, it is anticipated that the CERN Large Hadron Collider (LHC),
a $pp$ collider with centre of mass energy $\sqrt{s}=14$~TeV,
will reveal a new level of understanding of the fundamental interactions
when it starts to explore the TeV energy regime. For a number of reasons, including 
the quadratic sensitivity of the Higgs boson mass to radiative corrections, it
is generally believed that the Standard Model (SM) is a low energy effective limit
of a more fundamental theory and numerous extensions of the SM have been proposed.
Many of these extensions predict the existence of new 
neutral gauge bosons ($Z'$) and other $s$-channel resonances
\cite{Hewett:1988xc,Langacker:2008yv,Rizzo:2006nw,Leike:1998wr,Cvetic:1995zs}.
If a kinematically accessible $Z'$ exists, 
it is expected to be discovered very early in the LHC program.  
Once such an object is discovered, the 
immediate task would be to measure its properties and identify its origins. This is a difficult
task and there is a vast literature on $Z'$ observables and analysis techniques. 

A key ingredient in determining the nature of a new resonance is to measure it's couplings
to fermions.  The $Z'$ couplings to leptons can be measured using three observables: the 
cross section to leptons, the forward backward asymmetry, $A_{FB}$, and the width, $\Gamma_{Z'}$
\cite{Langacker:1984dc}.
For quarks, studies have shown that rapidity distributions 
can be used to separate $u$-quark couplings
from $d$-quark couplings \cite{delAguila:1993ym,Dittmar:2003ir}.  
However, these analyses are statistical in nature so there will always
be contributions from the other type of quark.  
In contrast, the ability to identify $b$- and $t$-quarks in the final state
can be a powerful tool to measure quark couplings 
that can be used to distinguish between models that give rise to $Z'$ bosons.  

Previous studies have pointed out that third generation fermions, top quarks in particular,
can be used to search for extra gauge bosons 
\cite{Harris:1999ya,Lynch:2000md,Agashe:2006hk,Khramov:2007ev,Lillie:2007yh,Baur:2008uv,Barger:2006hm}
and to distinguish between models \cite{Baur:2008uv,Frederix:2007gi}.
While some have noted the possibility 
of using third generation $t$- and $b$-quarks to distinguish between models of extra 
neutral gauge bosons \cite{Mohapatra:1992tc,Rizzo:1998ut,Rizzo:2006nw,Barger:2006hm}, 
this subject has not been fully explored.
The ability to identify heavy quark flavours
offers the unique opportunity to measure individual quark 
couplings that is not possible for light quarks. 
In what follows, we describe a method of
using $b$- and $t$-quark final states to distinguish between models of 
new physics that predict extra neutral gauge bosons 
\cite{Mohapatra:1992tc,martin}.  
The primary challenges in these measurements 
will be the identification efficiencies for top and bottom quarks needed
to make statistically meaningful measurements and 
the  discrimination of the $t$'s and $b$'s coming from $Z'$ decays 
from SM QCD backgrounds.


To distinguish between models, we propose to use the cross sections 
$\sigma(pp\to Z'  \to b\bar{b})$ and $\sigma(pp\to Z' \to t\bar{t})$, 
as described by the Drell-Yan cross section with the addition of a $Z'$
\cite{Langacker:1984dc,Barger:1980ix} at the LHC. 
We computed the cross sections using Monte-Carlo phase space integration
with weighted events,
imposing a rapidity cut on the final state particles of $|\eta|<2.5$
to take into account detector acceptances.
We also included $p_T $ and invariant mass
distribution cuts with values chosen to reduce QCD backgrounds as described below. 
In our numerical results we take $\alpha=1/128.9$, $\sin^2\theta_w=0.231$, $M_Z=91.188$~GeV, 
$\Gamma_Z=2.495$~GeV and $m_t=172.5$~GeV \cite{Yao:2006px}.
We use the CTEQ6M parton distribution functions \cite{Pumplin:2002vw} 
and included a K-factor to account for NLO QCD corrections \cite{KubarAndre:1978uy}  
while NNLO are not numerically important
to our results \cite{Melnikov:2006kv,Anastasiou:2003ds}.  
Final state QED radiation effects are important
\cite{Baur:2001ze} 
but require a detailed 
detector level simulation that is beyond the scope of the present analysis.
The $Z'$ widths only include decays to standard model fermions. NLO QCD 
and electroweak radiative corrections
were included in the width calculations \cite{Kataev:1992dg}.

An important challenge for this analysis will be to achieve sufficiently high 
$b$- and $t$-quark
identification efficiencies to provide 
the statistics needed to distinguish between models.
The ATLAS and CMS 
collaborations have worked hard at estimating these values but experience with real
data will be required to obtain reliable values.
We therefore present results for two sets of values, distilled from the literature,
that we expect to bound the values that will eventually be achieved by the LHC collaborations.
Once the LHC experiments start to collect data, these values should be refined as
experimeters gain experience and a better understanding of their detectors.

For $b$ identification efficiency,
the ATLAS TDR gives a value of  $\epsilon_b= 60\%$ for low luminosity running and 50\% for high 
luminosity running with 100 to 1 rejection against light and $c$-jets \cite{:1999fq}.  
We will use the latter 
value which is appropriate to the high luminosities we assume.
The rejection of fakes arising from light and $c$-jets 
can be improved considerably by requiring that both the $b$ and $\bar{b}$ are seen.  We 
therefore consider two cases for tagging $b\bar{b}$ events; 50\% when only one $b$ is observed
and $\epsilon_{b\bar{b}}=25 $\% when both the $b$ and $\bar{b}$ are detected, 
independent of the dijet mass. Note that 
the $b\bar{b}$ detection efficiency is likely to be higher than simply using $\epsilon_b^2$.

The understanding of $t$-quark identification efficiencies is evolving.  
The top quark almost always decays into a $b$-quark and a $W^+$ boson ($t\to W^+ b$) with the $W$'s 
subsequently decaying either into two leptons ($e\nu_e$, $\mu\nu_\mu$ or $\tau\nu_\tau$) or into 
a light quark-antiquark pair ($u\bar{d}$, $c\bar{s}$) that in turn hadronizes.  
The single lepton
plus jets final state, where one $W$ decays leptonically and the other $W$ decays hadronically, 
$t\bar{t}\to WWb\bar{b}\to (l\nu)(jj)b\bar{b}$, has a BR$\sim 30\% $ of all $t\bar{t}$ events and
is generally viewed as giving the best signal-to-background ratio. 
With suitable kinematic cuts and including the
BR to $(l\nu)(jj)b\bar{b}$, a recent ATLAS study estimates $\epsilon_{t\bar{t}}\sim 4\%$
\cite{Khramov:2007ev}.
However, reconstructing the invariant mass of the $t\bar{t}$ system will reduce this number
\cite{Khramov:2007ev}.  
The ATLAS TDR is slightly more optimistic, claiming the efficiency for detecting 
a $M_{t\bar{t}}=2$~TeV resonance of about 5\% including the semileptonic mode BR
while a CMS simulation obtains the lower value of $\epsilon_{t\bar{t}}\sim 2\%$  \cite{D'hondt:2007aj}.
Baur and Orr \cite{Baur:2008uv,Baur:2007ck} found 
that the $t$-quark identification efficiencies for this channel can be improved by
using 2-jet and 3-jet final states with $b$-tags.  
The fully hadronic modes have a combined BR $\sim  45\%$ so utilizing the hadronic modes has 
the potential of improving the $t\bar{t}$ identification efficiency significantly.
A method has been suggested to distinguish top jets from standard model backgrounds using 
substructure of the top jet \cite{Kaplan:2008ie,Thaler:2008ju}. 
Kaplan {\it et al.} \cite{Kaplan:2008ie} estimated that high $p_T$ dijets can be rejected with
an efficiency of $\sim 99.99\%$ while retaining $\sim 10\%$ of the $t\bar{t}$ pairs.
By combining the different top decay channels and identification strategies
it should be possible to increase the overall $t\bar{t}$ identification efficiency.
Given that the subject of $t$-quark identification
at the LHC continues to evolve,  we assume a wide range of values of $\epsilon_{t\bar{t}}$, taking 
1\% and 10\%  for the low and high efficiency scenarios respectively.

\begin{figure}[t]
\begin{center}
\centerline{\epsfig{file=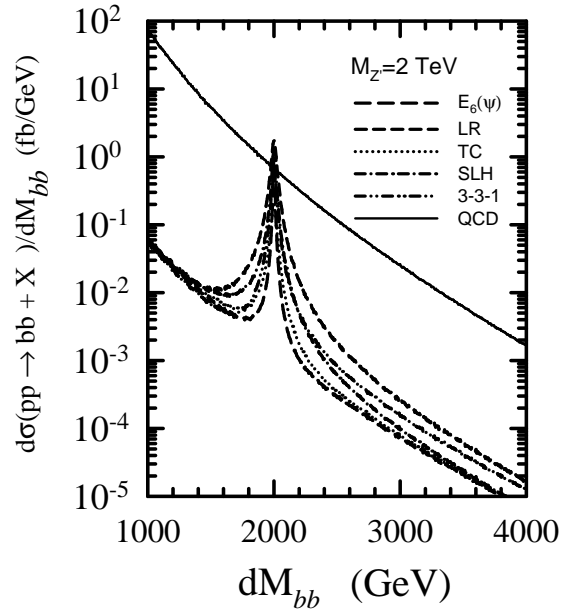,width=3.0in,clip=}}
\end{center}
\caption{Invariant mass distributions for the Drell-Yan process 
$pp\to b\bar{b}$  including a $Z'$ with mass $M_{Z'}=2$~TeV 
and the $b\bar{b}$ QCD backgrounds.  
The sets of curves correspond to 
$E_6(\psi)$ \cite{Hewett:1988xc}, 
Left Right symmetric $(g_R/g_L=1)$ \cite{Mohapatra:uf}, 
Simplest Little Higgs \cite{Schmaltz:2004de,ArkaniHamed:2001nc},
3-3-1 model \cite{Pisano:1991ee}, 
and TC models $(\tan\theta=0.577)$ \cite{Harris:1999ya,Hill:1991at}.
A kinematic cut of $P_T>50$~GeV was imposed on the $b$-quarks.
\label{Fig1}}
\end{figure}

Another challenge for making these measurements will be to 
distinguish the $Z'$ signal from the large SM QCD backgrounds.
The invariant mass distribution for $b\bar{b}$ final states is shown in
Fig.~\ref{Fig1} for 
the SM QCD background and the signal for a $Z'$ with a mass of 2~TeV for
several representative models.
The QCD backgrounds were calculated using the WHiZard package 
\cite{Kilian:2007gr}
with O'Mega matrix element generation \cite{Moretti:2001zz} and 
as an independent check we also calculated the QCD cross sections
using a simple Monte Carlo event generator with tree level matrix elements.
We use LO QCD cross sections in our background calculations. 
While it is known that higher
order QCD corrections can be substantial \cite{Baur:2007ck,Kidonakis:2003qe},
NLO corrections are highly dependent on the region of phase space being studied.  
As a crude estimate of the importance of NLO correction on our results, 
we rescaled the LO QCD backgrounds by a factor of 1.4
and found this to have little impact on our results.

\begin{figure}[t]
\begin{center}
\centerline{\epsfig{file=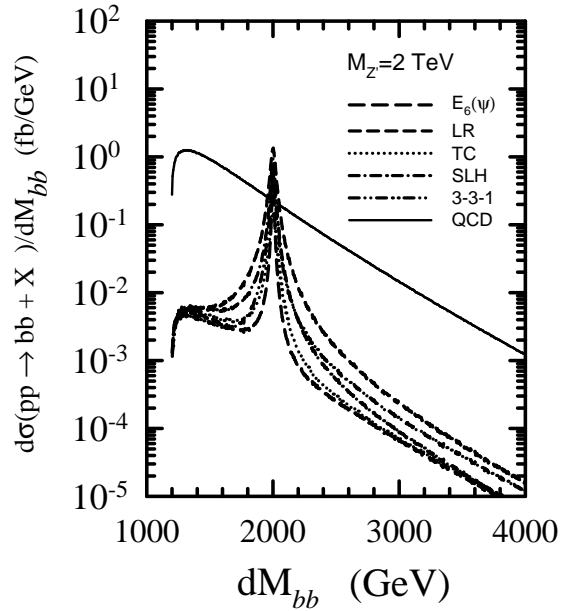,width=3.0in,clip=}}
\end{center}
\caption{Invariant mass distributions for the Drell-Yan process 
$pp\to b\bar{b}$  including a $Z'$ with mass $M_{Z'}=2$~TeV 
and the $b\bar{b}$ QCD backgrounds 
including a kinematic cut of $P_T>0.3 M_{Z'}$ on the $b$-quarks.  
\label{Fig2}}
\end{figure}

The $p_T$ distributions are quite different for the signal and backgrounds 
with quarks coming from $Z'$ decays having a much harder distribution than the
background events.  
The background can be reduced considerably by imposing a 
transverse momentum cut on the reconstructed final state $t$ and $b$'s
at some expense to the signal.  
The $p_T$ cut was varied and it was found that the optimum cut is approximately
$p_{T_Q}\geq 0.3 M_{Z'}$ which reduces the background significantly compared to
the signal.
A stronger
cut improves the signal to background ratio but decreases the total signal and therefore 
increases the statistical uncertainty. 
The invariant mass distribution for the signal and background are shown in
Fig.~\ref{Fig2}  after applying the cut.  

The QCD backgrounds can be further reduced by constraining the 
invariant mass of the final state fermions to 
$|M_{f\bar{f}}-M_{Z'} | \leq 2.5\Gamma_{Z'}$.
The window was chosen to balance the total signal against the 
signal to background ratio.  
We examined the model independent choice of 
$|M_{f\bar{f}}-M_{Z'} | \leq 0.07 M_{Z'}$,
but found that our results were not very sensitive to the precise choice
of $M_{f\bar{f}}$ window.

Fakes from gluon-, light quark-, and $c$-jets are potentially problematic but there is a
tradeoff between heavy quark identification efficiencies and mistagging that requires detailed
detector simulations.  Likewise, we defer detector resolution effects to more detailed 
future studies.
Other non-QCD SM 
backgrounds include $Wb\bar{b}+jets$, $(Wb+W\bar{b})$, $W+jets$, {\it etc} final states.  
Baur and Orr 
have shown that these can be controlled by constraining the
cluster transverse mass and invariant mass of outgoing jets (and leptons) to be close to $m_t$
\cite{Baur:2008uv,Baur:2007ck}.


In addition to the QCD backgrounds and the question of heavy quark identification 
efficiencies, there are additional theoretical uncertainties in the cross sections;
higher order QCD and EW corrections to the 
cross sections, both initial and final state contributions, and uncertainties in the parton
distribution functions.  
We can reduce some of these uncertainties by using ratios of heavy quark production 
to $\mu^+\mu^-$ production; $R_{b/\mu}$ and $R_{t/\mu}$.  In particular,
these ratios nearly eliminate the uncertainties originating in the parton distribution functions.  The 
ratios are defined by
\begin{eqnarray}
R_{b/\mu} &\equiv &\frac{\sigma(pp \rightarrow Z' 
\rightarrow b\overline{b})}{\sigma(pp \rightarrow Z' 
\rightarrow \mu^{+}\mu^{-})} \approx 
\frac{3K_{q}\left(g^{b 2}_{L} + g^{b 2}_{R}\right)}{\left(g^{\mu 2}_{L} + g^{\mu 2}_{R}\right)}
\label{eqn1} \\
R_{t/\mu} &\equiv &\frac{\sigma(pp \rightarrow Z' 
\rightarrow t\overline{t})}{\sigma(pp \rightarrow Z' 
\rightarrow \mu^{+}\mu^{-})} \approx 
\frac{3K_{q}\left(g^{t 2}_{L} + g^{t 2}_{R}\right)}{\left(g^{\mu 2}_{L} + g^{\mu 2}_{R}\right)}\; ,
\label{eqn2}
\end{eqnarray}
where $K_{q}$ is a constant depending on the QCD and EW correction factors, and the 
factor of 3 is due to summation over color final states. 
Each of these ratios depends on only four couplings from each model.
An analysis based on the 
location of a measured $Z'$ in the $R_{b/\mu}-R_{t/\mu}$ parameter space 
provides a means of distinguishing between models.

We assume that a $Z'$ has been discovered and it's mass and width measured 
\cite{Langacker:1984dc,Barger:1980ix}
so that the appropriate $M_{Q\bar{Q}}$ cuts described above can be applied.
It is expected that a $Z'$ with $M_{Z'} \leq 2$~TeV can be
discovered early in the LHC program with approximately 1-10~fb$^{-1}$ of integrated luminosity
depending on the specific model.

To obtain our results we calculate the expected number of events and statistical 
error for signal plus 
background for a given integrated
luminosity and particle identification efficiencies, $\epsilon_{\mu^+\mu^-}$,  
$\epsilon_{b\bar{b}}$, and $\epsilon_{t\bar{t}}$.  
The expected number of SM QCD and electroweak events were similarly calculated 
and subtracted from the signal plus background events to give the predicted number of 
signal events.   From these intermediate results we obtained the ratios given in equations
\ref{eqn1} and \ref{eqn2} with the errors calculated in the usual way by including both signal and
background.  We did not include uncertainties coming from luminosity and identification efficiencies.
In the latter case there is simply too big a range to include in an error, rather we show results
for the two cases discussed above.

\begin{figure}[t]
\begin{center}
\centerline{\epsfig{file=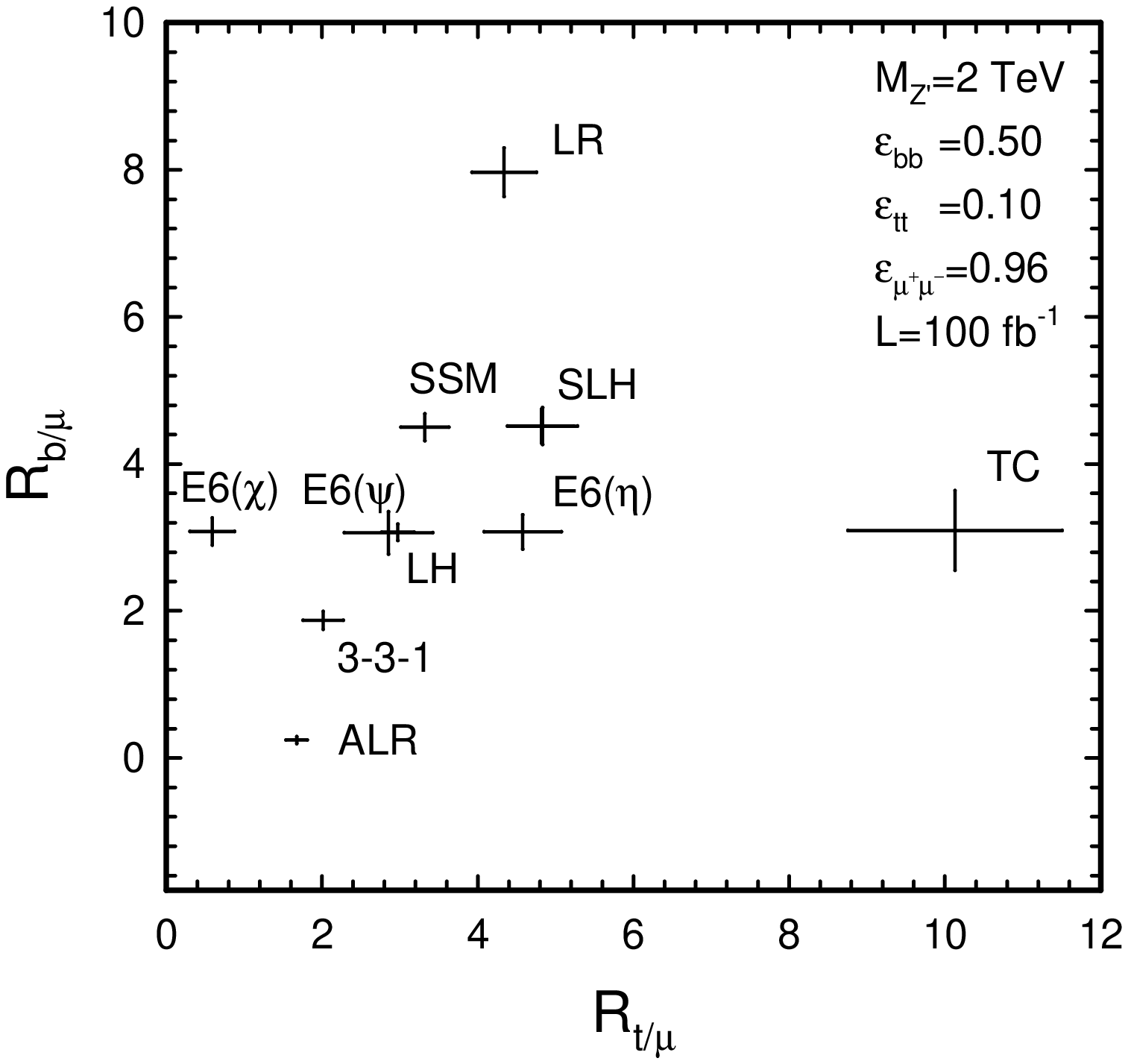,width=3.0in,clip=}}
\centerline{\epsfig{file=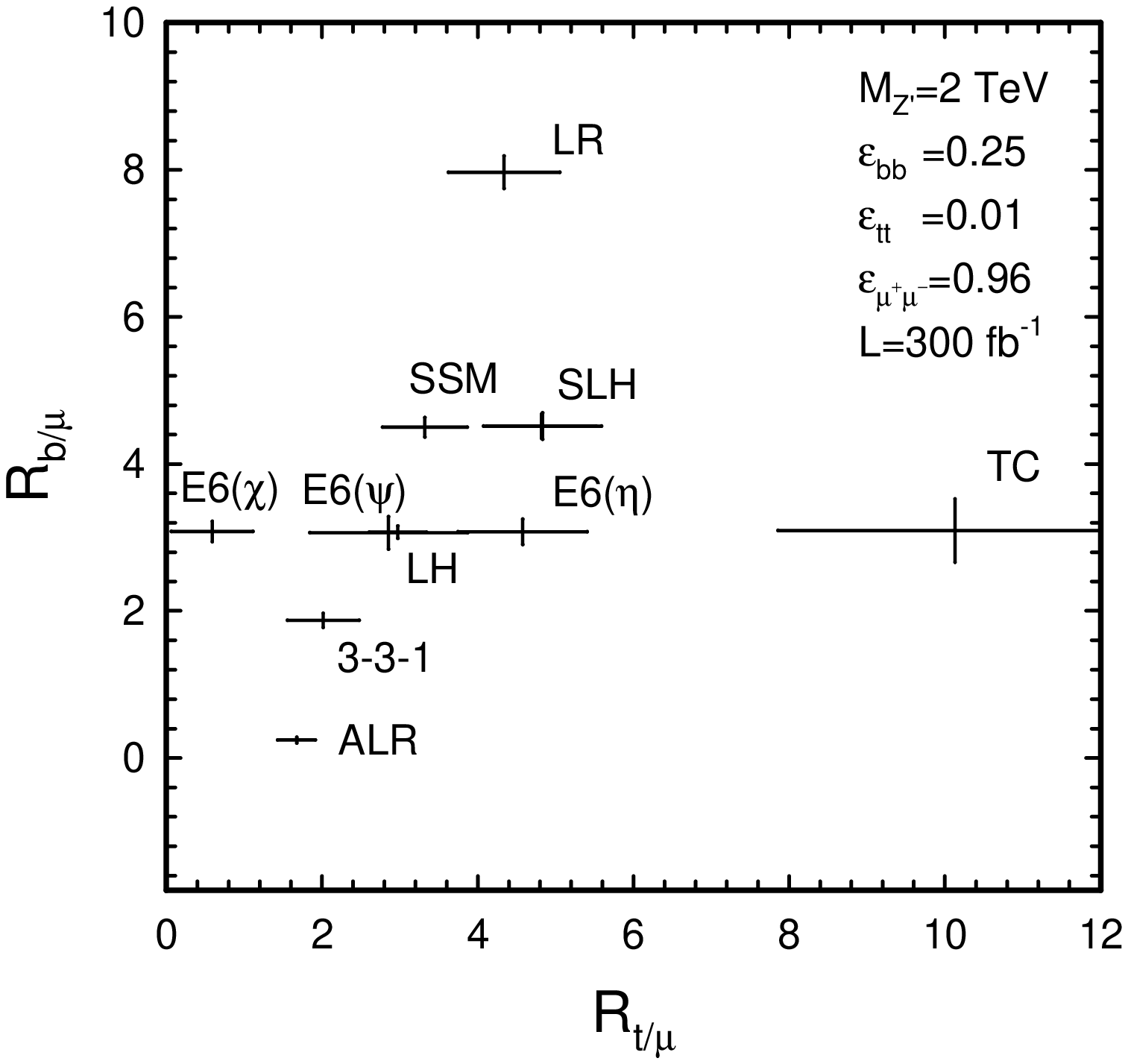,width=3.0in,clip=}}
\end{center}
\caption{$R_{b/\mu}$ versus $R_{t/\mu}$ for $M_{Z'}=2$~TeV for the 
$E_6(\chi)$, $E_6(\psi)$, $E_6(\eta)$ \cite{Hewett:1988xc}.,
Left-Right Symmetric Model $(g_R/g_L=1)$ (LR) \cite{Mohapatra:uf}, 
Alternate Left-Right Model $(g_R/g_L=1)$ (ALR) \cite{Ma:1986we},
Simplist Little Higgs Model (SLH) \cite{Schmaltz:2004de,ArkaniHamed:2001nc},  
Littlest Higgs Model $(\cot\theta_H=1)$ (LH) \cite{ArkaniHamed:2001nc,ArkaniHamed:2002qy},
3-3-1 2U1D Model \cite{Pisano:1991ee},
TC - Topcolour $(\tan\theta=0.577)$ \cite{Harris:1999ya,Hill:1991at}.
The error bars are the
statistical errors based on the integrated luminosity shown in the figure.
\label{Fig6}}
\end{figure}

Our results for $R_{b/\mu}$ and $R_{t/\mu}$ are shown in Fig.~\ref{Fig6} for $M_{Z'}=2$~TeV. 
Fig.~\ref{Fig6}(a) shows results for the high 
fermion identification efficiency values with $1\sigma$ statistical errors based on an 
integrated luminosity of $L=100$~fb$^{-1}$.
The low $\epsilon_{f\bar{f}}$ case would 
require higher integrated luminosity to distinguish between models so in 
Fig.~\ref{Fig6}(b) we show statistical errors based on $L=300$~fb$^{-1}$.
The errors scale as $1/\sqrt{L}$ and very roughly like $1/\sqrt{\epsilon_{Q\bar{Q}}}$ so one
can estimate how the errors will change with different integrated luminosities and heavy quark 
identification efficiencies.

It is clear that most models can  be differentiated using heavy quark final states.
However some models  such as the $E_6(\psi)$ and $SU(3)\times U(1)$ anomaly free Little Higgs 
model give similar ratios so one would need additional input such as
leptonic observables to distinguish between them.  


In summary, we demonstrated that, in principle, the decay of a $Z'$ boson into
third generation quarks can be used to distinguish between models of physics beyond
the SM.  The main challenge would be to reduce the measurement errors sufficiently
to discriminate between models and make accurate measurements of the $b$- and $t$-quark 
couplings to the $Z'$.  The major unknown in the analysis is the detection efficiency of 
the $t$ and $b$-quarks.  To account for this we considered two scenarios, an optimistic,
high efficiency  
scenario using  larger values for $\epsilon_t$ and $\epsilon_b$ given in 
the literature, and a pessimistic, low efficiency
scenario which used more conservative values.  We expect
that the LHC experiments will attain values somewhere in between.
Given the promise of this approach, a more detailed detector level study to see the effects of 
detector resolution is warranted.

\acknowledgments

The authors thank P. Kalyniak, H. Logan, H. Hou, J. Reuter, 
T. Rizzo, and T. Schwartz for helpful discussions and communications.
This research was supported in part by
the Natural Sciences and Engineering Research Council of Canada.


\end{document}